# Longitudinal analysis of fetal MRI in patients with prenatal spina bifida repair

Kelly Payette[1], Ueli Moehrlen[2], Luca Mazzone[2], Nicole Ochsenbein-Kölble[2], Ruth Tuura[1], Raimund Kottke[3], Martin Meuli[2], Andras Jakab[1]

[1] Center for MR-Research, University Children's Hospital Zurich, Zurich, Switzerland
`kelly.payette@kispi.uzh.ch`
[2] Fetal Surgery and Prenatal Consultation, University Children's Hospital Zurich, Zurich Switzerland
[3] Diagnostic Imaging and Intervention, University Children's Hospital Zurich, Zurich Switzerland

**Abstract.** Open spina bifida (SB) is one of the most common congenital defects and can lead to impaired brain development. Emerging fetal surgery methods have shown considerable success in the treatment of patients with this severe anomaly. Afterwards, alterations in the brain development of these fetuses have been observed. Currently no longitudinal studies exist to show the effect of fetal surgery on brain development. In this work, we present a fetal MRI neuroimaging analysis pipeline for fetuses with SB, including automated fetal ventricle segmentation and deformation-based morphometry, and demonstrate its applicability with an analysis of ventricle enlargement in fetuses with SB. Using a robust super-resolution algorithm, we reconstructed fetal brains at both pre-operative and post-operative time points and trained a U-Net CNN in order to automatically segment the ventricles. We investigated the change of ventricle shape post-operatively, and the impacts of lesion size, type, and GA at operation on the change in ventricle shape. No impact was found, except for moderately larger ventriculomegaly progression in myeloschisis patients. Prenatal ventricle volume growth was also investigated. Our method allows for the quantification of longitudinal morphological changes to fully quantify the impact of prenatal SB repair and could be applied to predict postnatal outcomes.

## 1    Introduction

Neural tube defects are some of the most common types of congenital defects and can lead to long-term physical and cognitive disabilities as well as social and psychological issues [1]. Myelomeningocele (MMC) and myeloschisis (MS) are open dysraphic neural tube defects, more commonly referred to as spina bifida (SB). They occur when the neural tube does not neurulate properly during development. Subsequently, the exposed part of the spinal cord suffers progressive damage during gestation. Advances in fetal magnetic resonance imaging (MRI) have better enabled identification and evaluation of SB at an early gestational age. The fetus can then undergo *in utero* repair of the spinal lesion in which the anatomy is reconstructed, protecting the spinal cord from any further damage during gestation and birth [2], [3]. The pro-cedure often leads to changes in brain structure, such as the reduction of the hindbrain herniation and the restoration of intracranial cerebrospinal fluid (CSF) space [4]. Reliability of the measurements of the posterior fossa and brain stem have been studied in SB using low resolution images, and while the posterior fossa measurements were reliable between observers, the brain stem measurements were not [5]. An early longitudinal MRI imaging analysis of the brain in SB fetuses would allow the visualization and objective quantification of changes occurring in the brain related to the pathology, and the impact of the surgical intervention on the developmental trajectory of the central nervous system.

Many advances have been made in the analysis of fetal MRI, especially in the creation of high resolution 3D volumes [6]–[10]. However, the analysis of these high-resolution volumes and their usage in longitudinal studies has not been explored in detail. In addition, the automatic segmentation of 3D volumes of pathological fetal brains and the usage of deformation-based morphometry (DBM) on fetal images to investigate brain growth of fetuses with SB have not been explored.

Automatic segmentation of fetal MRI has been explored in both pathological and healthy brains but has primarily focused on segmenting the fetal brain within the maternal tissue [10]–[14]. The segmentation of fetal brain tissues has been investigated to a lesser extent [12], [15]–[17]. However, many of these methods are atlas-based or rely on existing priors, which do not exist for fetal SB brains. Therefore, these methods are not yet applicable.

Another category of automatic segmentation methods that do not rely on atlases are convolutional neural networks (CNNs). One popular type of CNN for the automatic segmentation of medical images is the U-Net [18], which has been used for the localization and segmentation of the fetal brain within the overall fetal image [14] and has been tested on the segmentation of a normal fetal brain [19]. We propose to use the same network to segment brain tissues within a high-resolution 3D reconstructed volume of a fetal SB brain, both pre-operatively and post-operatively. These segmentations can then be used to perform a detailed longitudinal volumetric study of the brain growth and development of fetuses with SB.

In addition to a purely volumetric analysis, we propose the usage of DBM to explore changes in brain growth rates of the SB fetus through looking at deformation field for each subject between the pre-operative and post-operative period. This could potentially provide insight into regional changes in fetal brain shape, which isn't given just by looking at the volumes alone.

We propose a pipeline of fetal MRI image analysis, from fetal brain extraction and super-resolution reconstruction to automatic tissue segmentation of fetal SB brains and the quantification of structural brain growth in pathological brains using DBM. Due to its importance in SB and other congenital fetal pathologies, we demonstrate the applicability of a U-Net CNN to segment normal and dilated ventricles for longitudinal and volumetric analysis.

## 2    Methods

### 2.1    Dataset

Prenatal surgical repair of the MMC or MS lesion was carried out in 93 subjects at the Zurich Center for Fetal Diagnosis and Therapy between March 2012 and June 2019, with both pre- and post-operative MRIs available. Multiple MRI scans of the brain were acquired at each time point on 1.5T and 3T clinical GE whole-body scanners using a T2-weighted single-shot fast spin echo sequence (SSFSE), with an in-plane resolution of 0.5x0.5mm and a slice thickness of 3-5mm. The average gestational age in weeks (GA) of the subjects at the pre-operative scans, open fetal surgery, and post-operative scans were 23.2±1.5 GA, 25.0±0.8 GA, and 27.7±1.2 GA, respectively. The length and width of the overall lesion size was measured and recorded intra-operatively for each patient. The MRI images of 15 non-pathological fetal brains were acquired in the same manner as the fetal SB brains. These fetuses were scanned to confirm or rule out suspected abnormalities and were determined to have unaffected brain development.

### 2.2    Brain Extraction and Super-Resolution Reconstruction

We reoriented and masked the fetal brains in each of the individual low-resolution (LR) scans for each subject using a semi-automated atlas-based custom MeVisLab module. A super-resolution (SR) algorithm was then applied to each stack of images (comprising of between 3 and 14 LR scans, with at least one scan in each orientation: axial, sagittal, coronal) for each subject at both the pre-operative and post-operative time points, creating a 3D SR volume of brain morphology [9]. See Fig. 1a for an overview of the processing steps for each subject. The quality of the SR reconstructions was then reviewed. Of the original 93 cases, 44 cases had both high quality pre- and post- operative images, while the remaining cases were excluded due to excess fetal movement present in the LR scans, causing movement artefact in the images. This results in a sub-optimal SR reconstruction in either the pre- or post-operative time point (see Fig. 2 for an example of SR reconstructions). The majority of poor-quality SR reconstructions were at the pre-operative time point (36 cases), due to the relatively small size of the fetus allowing for increased motion. SR reconstructions of the 15 fetuses with normal brain development were performed in the same manor.

### 2.3    Automatic Ventricle Segmentation using U-Net

46 high quality SR volumes (15 normally developing fetal brains, 16 post-operative SB fetal brains, and 15 pre-operative SB fetal brains; the SB fetal brains were chosen from the larger 93 case sample set) were manually segmented into the following tissue classes: white matter, grey matter, ventricles, cerebellum, brain stem, CSF. More cases were not segmented due to the time-consuming nature of this task. Fetal brains from all three categories were chosen in order to create as general of a network as possible. The ventricle label was isolated from the overall segmentation in order to train the neural network. 43 ventricle cases were used as the training and testing data with an 80/20 split, with small amounts of data augmentation. The U-Net used in [14] was used and modified, adding batch normalization as well as changing the learning rate to 10E-5, and re-trained with the ventricle labels. The neural network was trained for 100 epochs. The additional 3 cases (one normally developing fetal brain, one pre-operative SB, and one post-operative SB) were preserved as an independent validation set (see Fig. 1b). A larger validation set would be desirable, however due to the relatively small training set we decided to maximize the cases used in the training in order to improve the performance of the network. Once the network was trained, the ventricles of the 46 post-operative and 46 pre-operative SR volumes were segmented using the network.

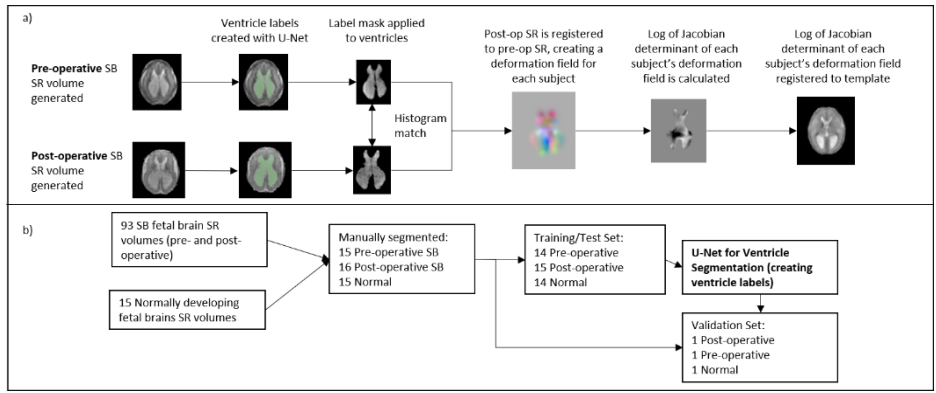

**Fig. 1.** a) Flowchart of processing steps at a subject level. For each subject, the post-operative ventricles are registered to the pre-operative ventricles, creating a deformation field. The log of the determinant of the Jacobian is calculated for each deformation field, which is then registered to the custom template. b) Flowchart of processing steps for training and validating the U-Net for ventricle segmentation. One case was kept out from each group (pre- and post-operative, and normally developing) for validation.

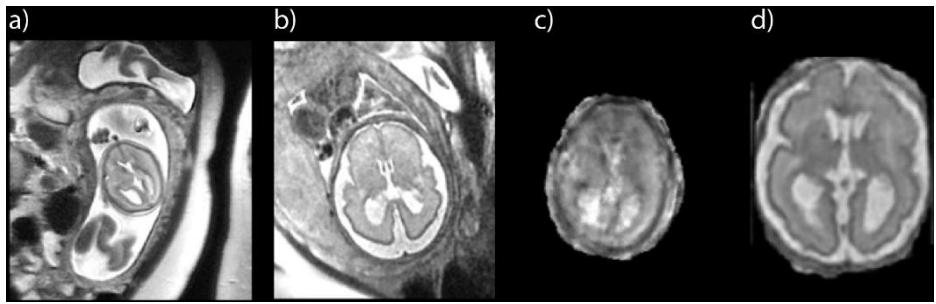

**Fig. 2.** Example SB SR reconstructions where the pre-operative reconstruction was poor (4 LR scans used, 21.6GA) and the post-operative reconstruction was successful (5 LR scans used, 27.0GA). a) example pre-operative LR scan, b) example post-operative LR scan, c) pre-operative SR volume, d) post-operative SR volume

## 2.4 Deformation-based Morphometry and Atlas Creation

The segmented ventricles of the pre-operative SR volume was resampled to 0.5mm and histogram-matched to the respective resampled post-operative SR volume [20]. Ventricle enlargement between the two time points was determined for each subject using DBM. The post-operative image was registered to the pre-operative image using a series of rigid, affine, and SyN registrations, creating a deformation field representing the ventricle enlargement. This field was then transformed into the same space, using a custom-made atlas of the post-operative SB brain. The custom atlas was created using the ANTS template reconstruction software [21], [22]. The log of the Jacobian determinant map of the deformation between the two images was calculated, and then scaled to create a daily Jacobian, as the difference between the two imaging time points was variable. The registrations and Jacobian calculations were performed using ANTs [23]. The transformed ventricle enlargement maps were then combined into a single 4D volume. In addition, the volumes of the ventricles were determined using the label maps.

## 2.5 Statistical Analysis

FSL's randomise was used to perform a statistical analysis of the ventricle enlargement maps using a general linear model with threshold free cluster enhancement [24]. We looked at the effect on ventricle enlargement of the area of the lesion, lesion type (MMC or MS), lesion location, and gestational age at operation. Gestational age at both the pre-operative and post-operative time points were controlled for, as well as gestational age at time of surgery.

## 3 Results

### 3.1 Ventricle Segmentation

Ventricle segmentation was assessed using the Dice overlap coefficient, comparing the segmentation from the U-Net with the ground truth (Table 1). The segmentations created by the U-Net can be seen in Fig. 3.

**Table 1.** Dice coefficients of ventricle segmentation validation of the U-Net CNN.

| Image | GA | Dice Coefficient | Sensitivity | Specificity |
|---|---|---|---|---|
| Pre-operative MS | 24.4 | 0.94 | 0.95 | 1.00 |
| Post-operative MMC | 30.1 | 0.91 | 0.91 | 1.00 |
| Normal | 23 | 0.91 | 0.90 | 1.00 |

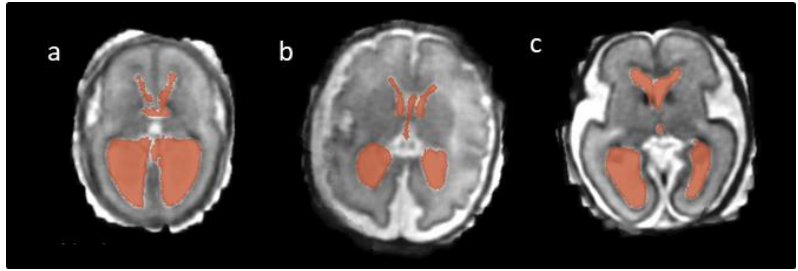

**Fig. 3.** Ventricle Segmentation with the U-Net of a) pre-operative MS brain (24.4GA); b) post-operative MMC brain (30.1GA); c) normally developing fetal brain (23GA)

### 3.2    Longitudinal Analysis

**Atlas Creation**

A custom template was generated for both pre-operative and post-operative fetal brains with the average GA at each time point. The pre-operative custom template was generated using 10 pre-operative SR volumes with an average GA of 23.1±0.3, and the post-operative template was generated using 20 post-operative SR volumes with an average GA of 27.3±0.3. There were fewer good quality SR volumes available for the creation of the pre-operative template. However, there are fewer features and less variation at younger gestational ages, therefore fewer subjects are required [25]. The atlases created can be seen in Fig. 4.

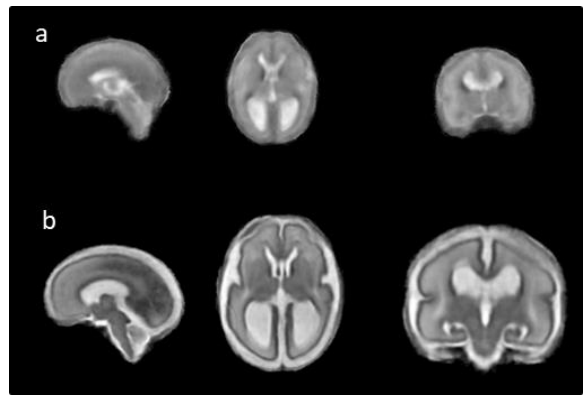

**Fig. 4.** Custom SB templates a) pre-operative, 23 GA; b) post-operative, 27 GA

**Volumetry**

From the segmented volumes, the volumes of each of the fetuses at both time points were determined and can be seen in Fig. 5 and Table 2. Two cases were excluded due to failure of the segmentation algorithm, mainly due to the quality of the SR reconstruction. The average daily volume growth within the ventricles was 449.8±285.9mm$^3$. All SB fetal ventricle volumes were determined based on segmentations created using the U-Net network and were reviewed for accuracy. In some cases where there were errors, minor adjustments were made to ensure that the wide range of ventricle volumes found reflected the pathology and not errors in segmentation. All normal fetal ventricle volumes were determined through manual segmentation and taken from the training set used for the U-Net.

**Table 2.** Ventricle volume overview for SB pre-operative, SB post-operative, and normal groups. The SB volumes were found using the U-Net, and the normal volumes from the manual segmentations performed for the U-Net training set.

| Dataset | n | Average GA | Average Volume (mm3) |
| --- | --- | --- | --- |
| Pre-operative SB | 44 | 23.5±1.4 | 8953±6345 |
| Post-operative SB | 44 | 27.5±1 | 20480±10922 |
| Normal | 15 | 27.9±3.5 | 7139±3063 |

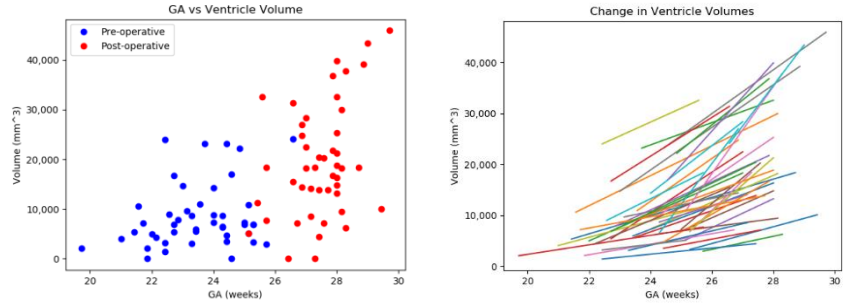

**Fig. 5.** Left: GA vs ventricle volume. Blue: pre-operative, Red: post-operative. Right: Change in ventricle volume for each fetus.

**DBM**

Deformation based morphometry was performed on the segmented ventricles for each case and registered to the custom template. Mean ventricle enlargement can be seen in Fig. 6. As expected, ventricle enlargement occurs in all sections of the ventricles. However, some asymmetry in the ventricles is present, most likely due to the underlying pathology of the SB patients. The SR volumes used to create the atlas were a subset of the overall population, and there was potentially more asymmetry in the overall population than in the atlas subgroup.

In addition, a regression analysis using threshold free cluster enhancement was performed investigating the impact of lesion size, lesion type, and GA at operation on ventricular enlargement. Lesion size, lesion type. and GA at operation had no impact on the ventricular enlargement.

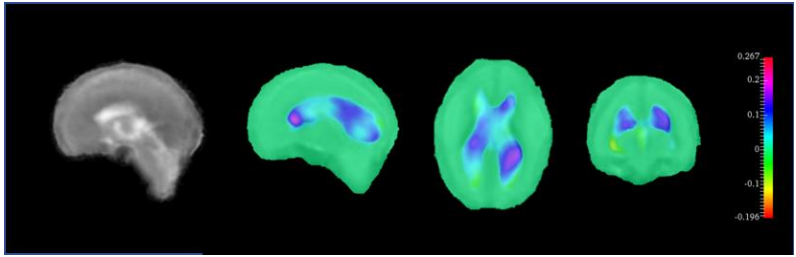

**Fig. 6.** Left: Custom pre-operative atlas; Right: Mean fetal ventricle enlargement (log of the Jacobian determinant of the deformation field) between the pre-operative and post-operative time points overlaid on the custom pre-operative atlas.

## 4     Discussion and Conclusion

In this work, we have presented a fetal MRI image analysis pipeline for fetuses with SB, including automated fetal ventricle segmentation. We demonstrated its applicability with an analysis of ventricle enlargement in fetuses with SB. Using a robust super-resolution algorithm [9], we reconstructed fetal brains at both the pre-operative and post-operative time point, and then used 46 manual segmentations, applied data-augmentation, and trained a U-Net CNN in order to automatically segment the ventricles within the SB SR volume. We then used DBM to investigate the change in shape of ventricles of the SB fetuses post-operatively and the impact of lesion size, type, and GA at operation on the change in ventricle shape, all of which had no impact.

Overall, the quality of the automatic segmentation depended on the quality of the super-resolution reconstruction. As many subjects were excluded due to poor quality reconstructions, it would be beneficial to improve this step in order to reduce the number of subjects excluded.

The volumetric analysis of SR volumes using the U-Net is very promising and will be expanded to other brain tissues. Further data augmentation methods can potentially be applied in order to increase the training set without having to manually segment more volumes, as it is an incredibly time-consuming process prone to error.

The use of DBM as a method of quantifying brain growth is very dependent on the atlas creation and improved atlases would be beneficial. The pre-operative atlas is fuzzy when compared to standard atlases of healthy fetal brains. However, the quality of the atlas is directly related to the quality of the input volumes, which for the young pre-operative fetal brains are not as sharp as the post-operative brains. In addition, one cannot draw any results from only looking at the change in shape of the ventricles, as that does not consider all the other changes in the brain occurring. A more complete analysis would need to investigate all brain tissues.

In the future, we aim to expand the analysis of tissue brain growth of SB fetuses to other structures of the brain. In particular, we aim to explore the post-operative changes in the hindbrain herniation occurring in the context of Chiari II malformation. We aim to quantify the impact of fetal SB repair, and ideally to better predict the outcome of the surgery in order to aid prenatal parental guidance.

## Acknowledgements

Financial support was provided by the OPO Foundation, Anna Müller Grocholski Foundation, the Foundation for Research in Science and the Humanities at the University of Zurich, EMDO Foundation, Hasler Foundation, and the Forschungszentrum für das Kind Grant (FZK).

*Author's original version (pre-print).*